\documentclass[12pt]{article}
\usepackage{geometry}
\usepackage{authblk} 
\usepackage{lineno} 
\usepackage{microtype}
\usepackage{hyperref}
\usepackage{url}
\usepackage{amsbsy} 
\usepackage{amssymb}
\usepackage{amsmath}
\usepackage{flexisym}
\usepackage{graphicx}
\usepackage{caption}
\usepackage{subcaption}
\usepackage{natbib}
\usepackage{multirow}
\usepackage{arydshln}
\usepackage[usenames,dvipsnames]{color}

\date{ \today }

\begin{document}

\title{A Seismologically Consistent Surface Rupture Length Model for Unbounded and Width-Limited Events 
\large{ \textit{(Submited to Earthquake Spectra)} }
}

\author[1,2]{Grigorios Lavrentiadis\thanks{glavrentiadis@ucla.edu; glavrent@caltech.edu}} 
\author[3]{Yongfei Wang}
\author[4]{Norman A. Abrahamson}
\author[2]{Yousef Bozorgnia}
\author[3]{Christine Goulet}
\affil[1]{Mechanical and Civil Engineering Department,\protect\\ California Institute of Technology}          
\affil[2]{Natural Hazards Risk and Resiliency Research Center, \protect\\
          The Garrick Institute for Risk Sciences, \protect\\
          University of California, Los Angeles}
\affil[3]{Southern California Earthquake Center,\protect\\ University of Southern California}
\affil[4]{Department of Civil and Environmental Engineering,\protect\\ University of California, Berkeley}

\maketitle

\begin{abstract}
    A new surface-rupture-length ($SRL$) relationship as a function of magnitude ($\mathbf{M}$), fault thickness, and fault dip angle is presented in this paper. 
    The objective of this study is to model the change in scaling between unbounded and width-limited ruptures. 
    This is achieved through the use of seismological-theory-based relationships for the average displacement scaling and the aid of dynamic fault rupture simulations to constrain the rupture width scaling. 
    The empirical dataset used in the development of this relationship is composed of $123$ events ranging from $\mathbf{M}~5$ to $8.1$ and $SRL~1.1$ to $432~km$.
    The dynamic rupture simulations dataset includes $554$ events ranging from $\mathbf{M}~4.9$ to $8.2$ and $SRL~1$ to $655~km$. 
    For the average displacement ($\bar{D}$) scaling, three simple models and two composite models were evaluated.
    The simple average displacement models were: a square root of the rupture area ($\sqrt{A}$), a down-dip width ($W$), and a rupture length ($L$) proportional model. 
    The two composite models followed a $\sqrt{A}$ scaling for unbounded ruptures and transitioned to $W$ and $L$ scaling for width-limited events, respectively. 
    The empirical data favors a $\bar{D} \sim \sqrt{A}$ scaling for both unbounded and width-limited ruptures. 
    The proposed model exhibits better predictive performance compared to linear $\log(SLR)\sim\mathbf{M}$ type models, especially in the large magnitude range, which is dominated by width-limited events.
    A comparison with existing $SRL$ models shows consistent scaling at different magnitude ranges that is believed to be the result of the different magnitude ranges in the empirical dataset of the published relationships. 
\end{abstract}

\section{Introduction}

Surface-rupture fault displacement hazard analyses, both probabilistic and deterministic, require an estimate of the Surface Rupture Length ($SRL$) to compute the shape of the slip profile \citep{Youngs2003, Petersen2011, Moss2011, Lavrentiadis2019}.
Studies such as \cite{Wells1994} and \cite{Wells2015} proposed empirical models for the $SRL$ scaling using observations from past earthquakes.
These models are straightforward to develop and have good predictive performances within the range of data, but they may exhibit poor extrapolation for large events as the mechanisms that control the $SRL$ scaling in large earthquakes is not well understood due to the limited empirical data. 
Another approach for developing scaling relationships is based on the use of theoretical considerations and constraints. 
An example of such model is \cite{Leonard2010} who developed a set of equations that describes the scaling between seismic moment, rupture area, length, width, and average displacement by imposing a self-consistent constraint.
This type of model may have slightly larger aleatory variability at the center of the data but exhibit better extrapolation behavior due to the incorporation of seismological constraints.

The impact of the thickness of the seismogenic zone in the geometry of the rupture has been observed by previous studies, for instance, \cite{Hanks2002} who proposed a magnitude break in their $\log(A) \sim \mathbf{M}$ relationship at magnitude $\mathbf{M} = 6.71$.
However, to the authors' knowledge, a relationship between moment magnitude ($\mathbf{M}$) and $SRL$ that considers the effect of the finite seismogenic zone thickness does not exist. The recent development of extensively documented community-based fault displacement databases and advances in computer-based earthquake rupture simulation methods (both described below) motivated the development of the model presented herein, which explores this issue. 
The objective of this study is to develop a new $SRL$ relationship that captures the changes in scaling between unbounded and width-limited ruptures.
The impact of the change in scaling is expected to be more pronounced in active crustal regions (ACRs) with thin seismogenic zones, such as California, as opposed to stable-continental regions with thicker seismogenic zones, such as Australia or the northeastern North American region. 

\begin{figure}[htbp!]
    \centering
    \begin{subfigure}[t]{0.48\textwidth}
        \caption{}
        \includegraphics[height = 0.4\textwidth]{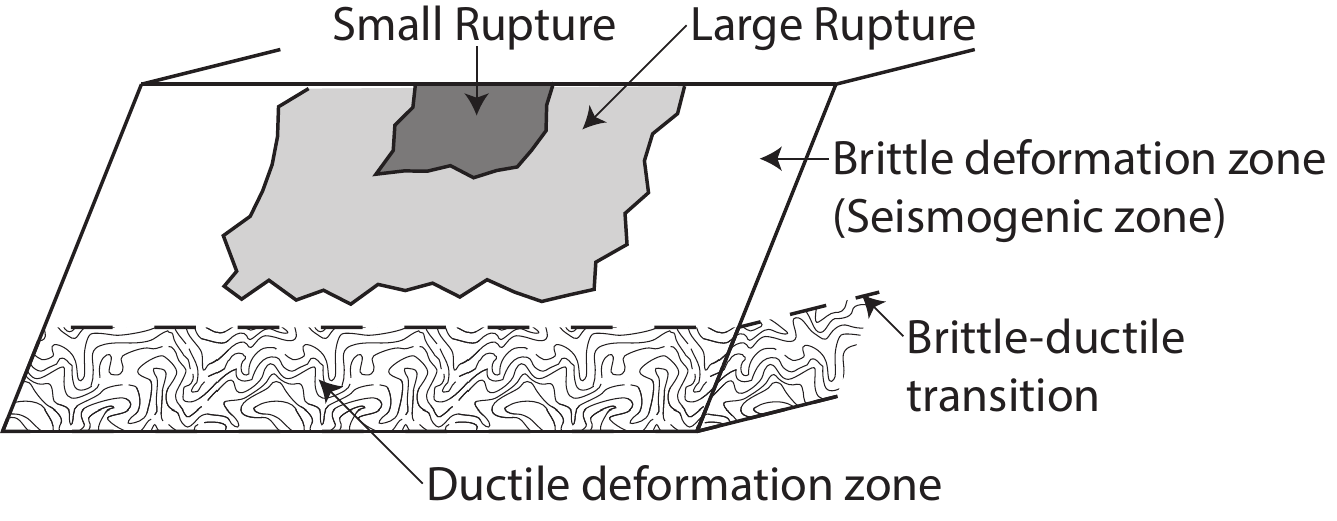}
    \end{subfigure}
    \begin{subfigure}[t]{0.48\textwidth}
        \caption{}
        \includegraphics[height = 0.4\textwidth]{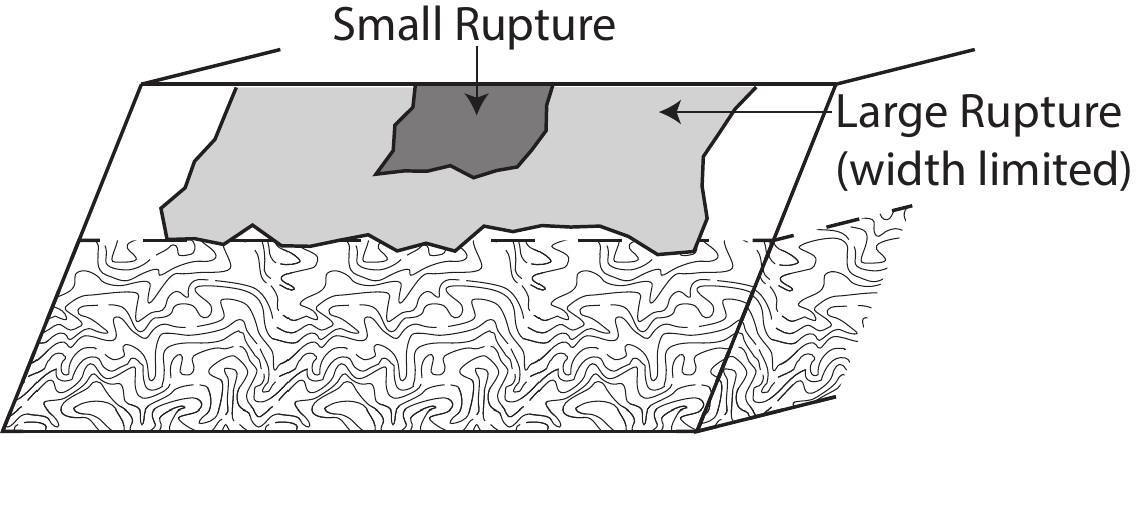}
    \end{subfigure}
    \caption{Sketches of rupture geometries for small (dark shading) and moderate-to-large (pale shading) events (a) in thick crust regions and (b) in think crust regions }
    \label{fig:sketch_rup}
\end{figure}

A sketch illustrating the two conditions is presented in Figure \ref{fig:sketch_rup}.
Assuming a similar static stress drop between the two regions, events of similar magnitude are expected to have similar rupture areas ($A$) (Equation \ref{eq:brune71} from \cite{Brune1971}).
Additionally, making the assumption that ruptures grow with similar aspect ratio scaling, small events that produce surface rupture, which do not reach the bottom of the seismogenic zone, are expected to have similar rupture geometries (rupture length and width), between the two regions.
However; for moderate-to-large events, the thinner seismogenic zone is expected to impede the growth of the rupture width at shallower depths, requiring the rupture length to grow faster in order to accommodate the same rupture area, which will result in a magnitude break and steeper scaling for $SLR \sim \mathbf{M}$ relationship proposed herein.

\begin{equation} \label{eq:brune71}
  \log(M_o) = 3/2 \log(A) + \log(\Delta \sigma) - 0.387
\end{equation}

\section{Data}

Two sets of data were used in developing the proposed model: an empirical dataset used for the $SRL$ scaling (\hyperref[sec:emp_data]{Subsection: Empirical Data}) and a numerical phyics-based simulation dataset used to constrain the rupture width scaling (\hyperref[sec:scec_data]{Subsection: SCEC Simulations}).
The empirical and numerical simulation datasets are included in the electronic supplement of this article.

\subsection{Empirical Data} \label{sec:emp_data}

The empirical datasets used in the model development include: (i) the Fault Displacement Hazard Initiative (FDHI) dataset \citep{Sarmiento2021}, (ii) additional events from \cite{Wells1994}, which they classified as reliable, and (iii) the surface rupture events from  \cite{Baize2020} which were not part of the FDHI dataset.
Events less than $\mathbf{M}~5$ were excluded from the final dataset, as well as the 1892 Laguna Salada, 1978 IzuOshima, and 1993 Killari earthquakes which were considered outliers.
Potential reasons for considering the previous events as outliers are: the Laguna Saldana earthquake ruptured in the 1800s but was mapped in the 2010s, so large parts of the rupture may have been irrecoverable, the 1978 Izu Oshima earthquake likely included an offshore segment that was not mapped in the FDHI dataset, and the 1993 Killari occurred in a stable continental region with a lot of deformation probably accommodated via folding/warping that was unmapped. 

\begin{figure}
    \centering
    \includegraphics[height=0.5\textwidth]{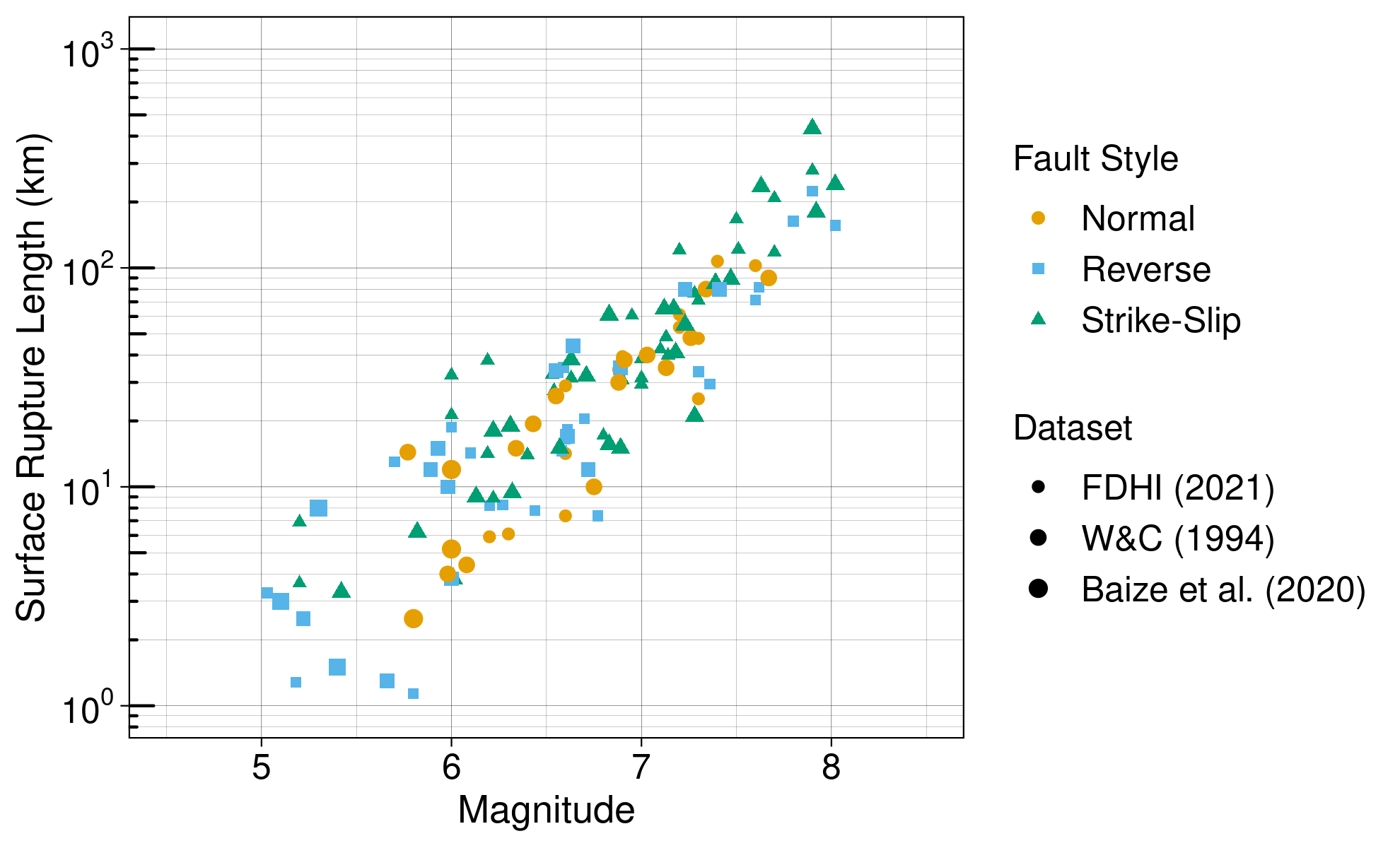}
    \caption{Comparison of Magnitude $\mathbf{M}$ - Surface Rupture Length ($SRL$) distribution of empirical datasets.}
    \label{fig:empirical_datasets}
\end{figure}

The $SRL$ values for the events from \cite{Wells1994} and \cite{Baize2020} were obtained directly from the corresponding studies, while for the events in the FDHI dataset, $SRL$ was estimated based on the length of the Event Coordinate System (ECS) defined in \cite{Lavrentiadis2022a}
Figure \ref{fig:empirical_datasets} shows the $SRL$ versus $\mathbf{M}$ distribution of the final empirical dataset. 
It is composed of $65$ events from FDHI, $52$ events from \cite{Wells1994}, and $6$ events form \cite{Baize2020}.
The earthquake $\mathbf{M}$ ranges from $5.0$ to $8.1$, while $SRL$ ranges from $1.1$ to $432~km$.  
It contains $65$ strike-slip, $26$ normal, and $32$ reverse earthquakes. 

In absence of fault-specific data, the fault width ($W_{lim}$) required to define the transition to a width-limited rupture, was estimated with Equation \ref{eq:fault_width} based on the thickness of the seismogenic zone ($D_{seis}$) and the fault dip angle ($\theta$).
$D_{seis}$ was based on the fault location as summarized in Table \ref{tab:seis_thick}, whereas $\theta$ was based on the style of faulting, as summarized in Table \ref{tab:fault_dip}. 

\begin{equation} \label{eq:fault_width}
    W_{lim} = \frac{D_{seis}}{\sin(\theta)}
\end{equation}

\begin{table}[htbp!]
    \caption{Seismogenic Thickness Values for Different Regions}
	\centering
	\label{tab:seis_thick}
	\begin{tabular}{|l|c|}
        \hline
        Region              & Seismogenic Thickness ($km$) \\ \hline 
        California	        &	15	\\ \hline
        Guatemala	        &	15	\\ \hline
        New Zealand	        &	15	\\ \hline
        Indonesia	        &	15	\\ \hline
        Japan	            &	17	\\ \hline
        Himalayan Region    &	30	\\ \hline
        Australia	        &	40	\\ \hline
        Other Regions       &	22	\\
        \hline
	 \end{tabular}
\end{table}

\begin{table}[htbp!]
	\caption{Dip Angle Values for Different Styles of Faulting}
	\centering
	\label{tab:fault_dip}
	\begin{tabular}{|l|c|}
        \hline
	    Style of Faulting              & Dip Angle ($deg$) \\
        \hline
	    Strike-slip &	90	\\ \hline
        Normal      &	60	\\ \hline
        Reverse     &   45	\\
        \hline
	 \end{tabular}
\end{table}

\subsection{Dynamic Rupture Simulations} \label{sec:scec_data}

\begin{figure}
    \centering
    \includegraphics[height=0.5\textwidth]{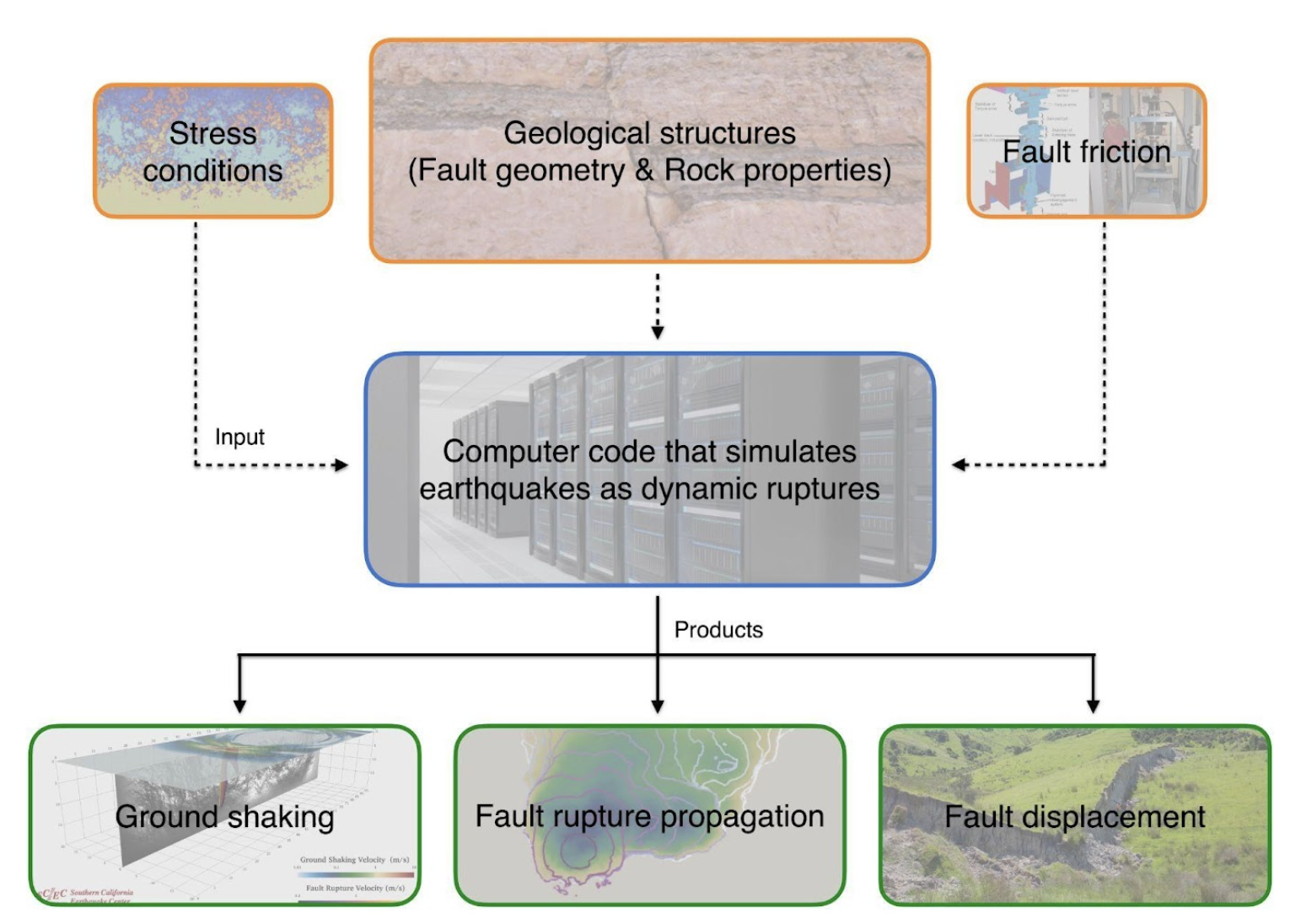}
    \caption{Key ingredients and schematic pipeline of a dynamic earthquake rupture simulation. Inputs include the initial stress conditions, the fault structure, the properties of the nearby rocks, and formulations that describe how the fault slips. A computer program numerically solves the resultant fault rupture propagation and wave propagation and outputs the ground shaking and fault displacements.}
    \label{fig:dynamics}
\end{figure}

The dataset used for constraining the rupture width scaling is based on physics-based fault displacement simulations conducted by the Southern California Earthquake Center (SCEC) as part of the FDHI efforts. 
The physics-based approach used in this study is referred to as the dynamic rupture model \citep{Harris2018}; it constructs spontaneously evolving earthquake ruptures under mechanical causative conditions (e.g., fault geometry, friction laws, stress conditions, and surrounding rock properties). Figure \ref{fig:dynamics} illustrates a simplified workflow of the dynamic rupture model. 
In contrast to the general theoretical considerations such as used by \cite{Leonard2010}, the dynamic rupture model employs a physically plausible parametric uncertainty (e.g., heterogeneous initial stress) to capture the displacement variability across the rupture plane due to the elasto-plasto-dynamic response to imposed stresses. The surface rupture length is then computed using numerical criteria. For the purpose of this paper, the length is represented by summing the length of fault segments with surface displacement larger than $1~cm$. 

The fault displacements were simulated by numerically solving the 3D elastoplastic spontaneous rupture propagation with the Support Operator Rupture Dynamics (SORD) code \citep{Wang2021,Wang2020,Wang2017}. This application is highly optimized and scalable on current cutting-edge supercomputers. This is a requirement since to capture a $\mathbf{M}$ range of $5$ - $8$ associated with rupture lengths of several $km$ to hundreds of $km$, the computational demands vary from seconds on tens of CPUs to hours on hundreds of thousands of CPUs. The simulations described here were performed on Theta at the Argonne Leadership Computing Facility and Frontera at the Texas Advanced Computing Center.

\begin{figure}
    \centering
    \includegraphics[height=0.5\textwidth]{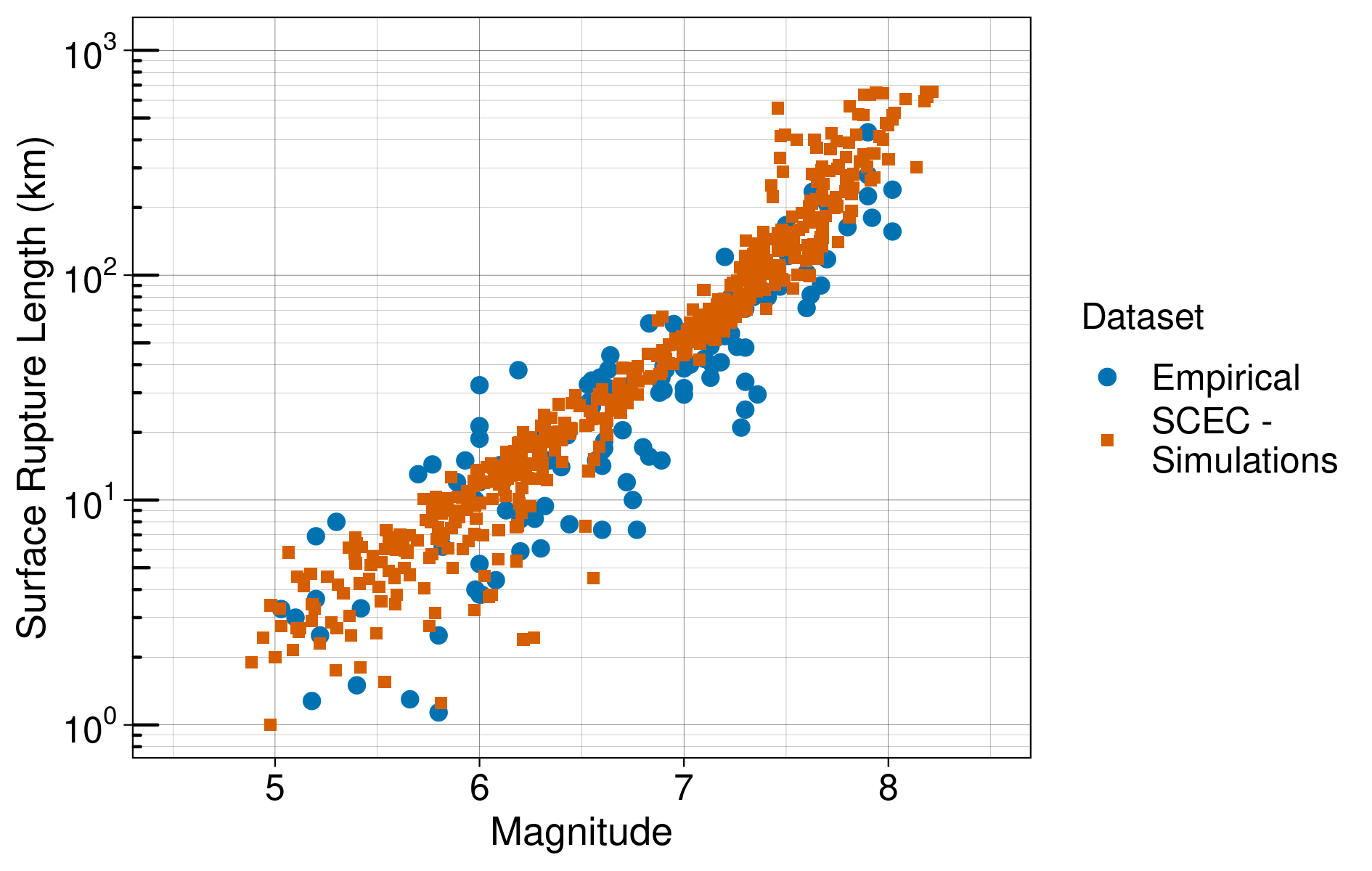}
    \caption{Comparison of Magnitude - Surface Rupture Length ($SRL$) distribution of empirical datasets and SCEC simulations.}
    \label{fig:empirical_scec_datasets}
\end{figure}

A series of dynamic fault rupture simulations under various model setups were therefore performed to understand and model the effect of fault width in the $SRL$ scaling. The general parameters were set based on the work of \citep{Wang2021}, with additional variations, as described below, all for vertical strike-slip faults. One set of simulations was performed in which the fault width was limited to $19km$, with two variations in the normal loading stresses pattern: fixed with depth in one case, and linearly increasing with depth in the other. The other set of simulations involved limiting the fault width to $15km$ while maintaining the other model parameters fixed. It is noted that although a width limit is provided for the dynamic rupture models ($15km$ and $19km$), it is not prescribed.  The resulting simulations output also includes the $\mathbf{M}$ which is not a-priori imposed, and the displacement field across the fault plane. In other words, the output result from spontaneously constrained in-situ boundary conditions and physical laws, which leads to the scatter visible in Figure \ref{fig:empirical_scec_datasets}. The simulations were continued to generate a wavefield and to provide surface ground motions, which were verified against existing ground motion models (GMMs), to ensure that the simulations are reasonable and technically defensible.  The detailed description of the dynamic rupture model can be found in \citep{Wang2022}.

In total, $554$ simulations were performed with magnitude ranging from $4.9$ to $8.2$ and $SRL$ ranging from $1$ to $655 km$
Figure \ref{fig:empirical_scec_datasets} compares the $\mathbf{M} - SRL$ scaling of the empirical and SCEC datasets. 
Overall, the empirical and simulation datasets appear consistent, supporting the use of the simulation datasets to inform the mean $SRL$ scaling in the empirical data. 
The aleatory variability of the SCEC dataset is expected to be less than that of the empirical data, as the goal of the simulations was to capture average scaling effects and not to fully represent the randomness and uncertainty in earthquake ruptures. 

\section{Model Development}

The derivation of the $SRL$ model is divided into two parts. 
In the first part, an average model for the rupture width ($W$) scaling of unbounded and width-limited ruptures based on the SCEC simulations is proposed.
In the second part, the $SRL$ model based on the empirical data is developed with the aid of the $W$ model to capture the transition between unbounded and width-limited ruptures. 

\subsection{Rupture Width Modeling} \label{sec:model_W}

\begin{figure}
    \centering
    \includegraphics[height=0.5\textwidth]{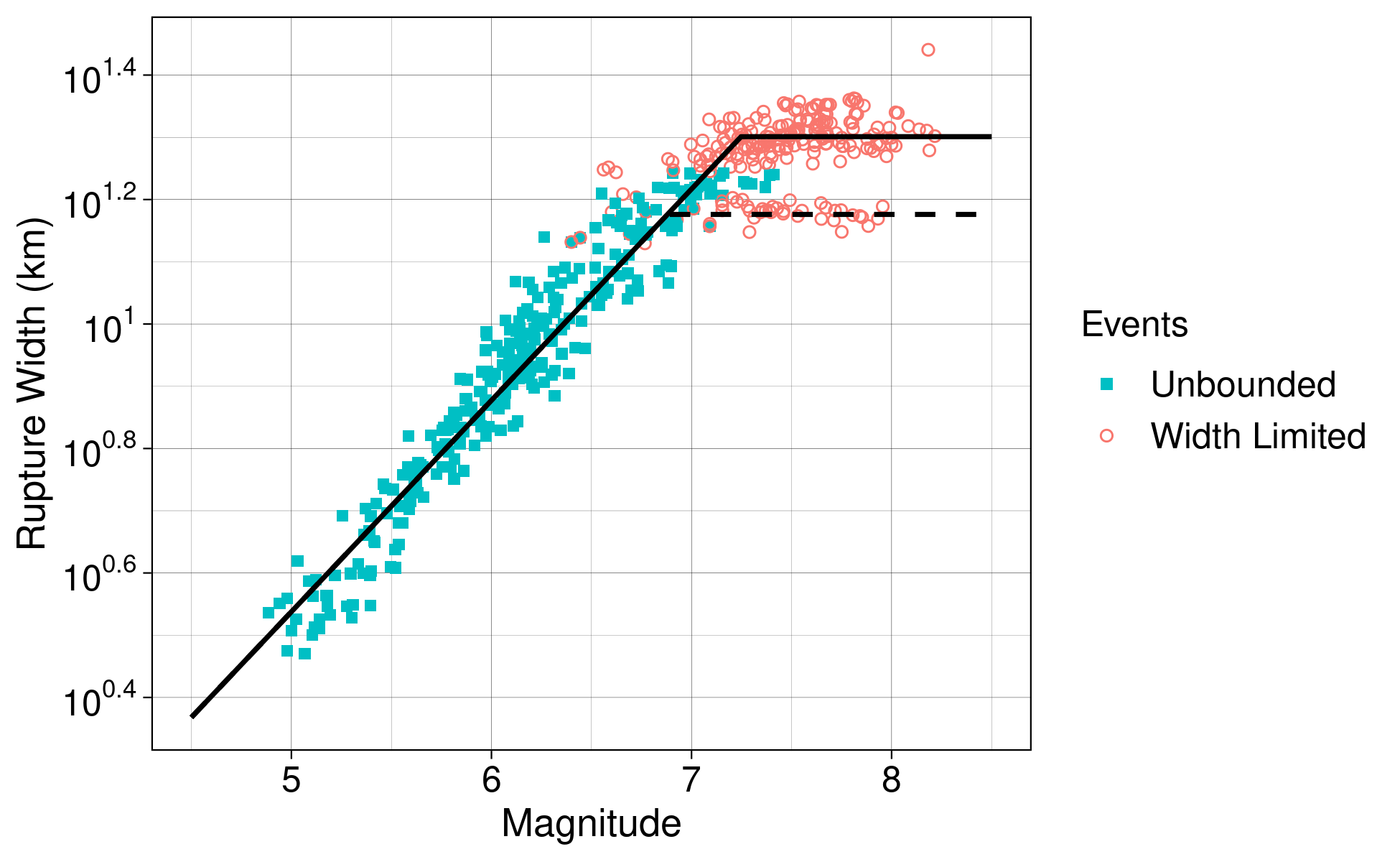}
    \caption{Comparison of Magnitude - Rupture Width model with SCEC simulations. The solid vertical line corresponds to a $19 km$ fault width, dashed vertical line corresponds to a $15 km$ fault width.}
    \label{fig:model_W_prediction}
\end{figure}

\begin{figure}
    \centering
    \includegraphics[height=0.5\textwidth]{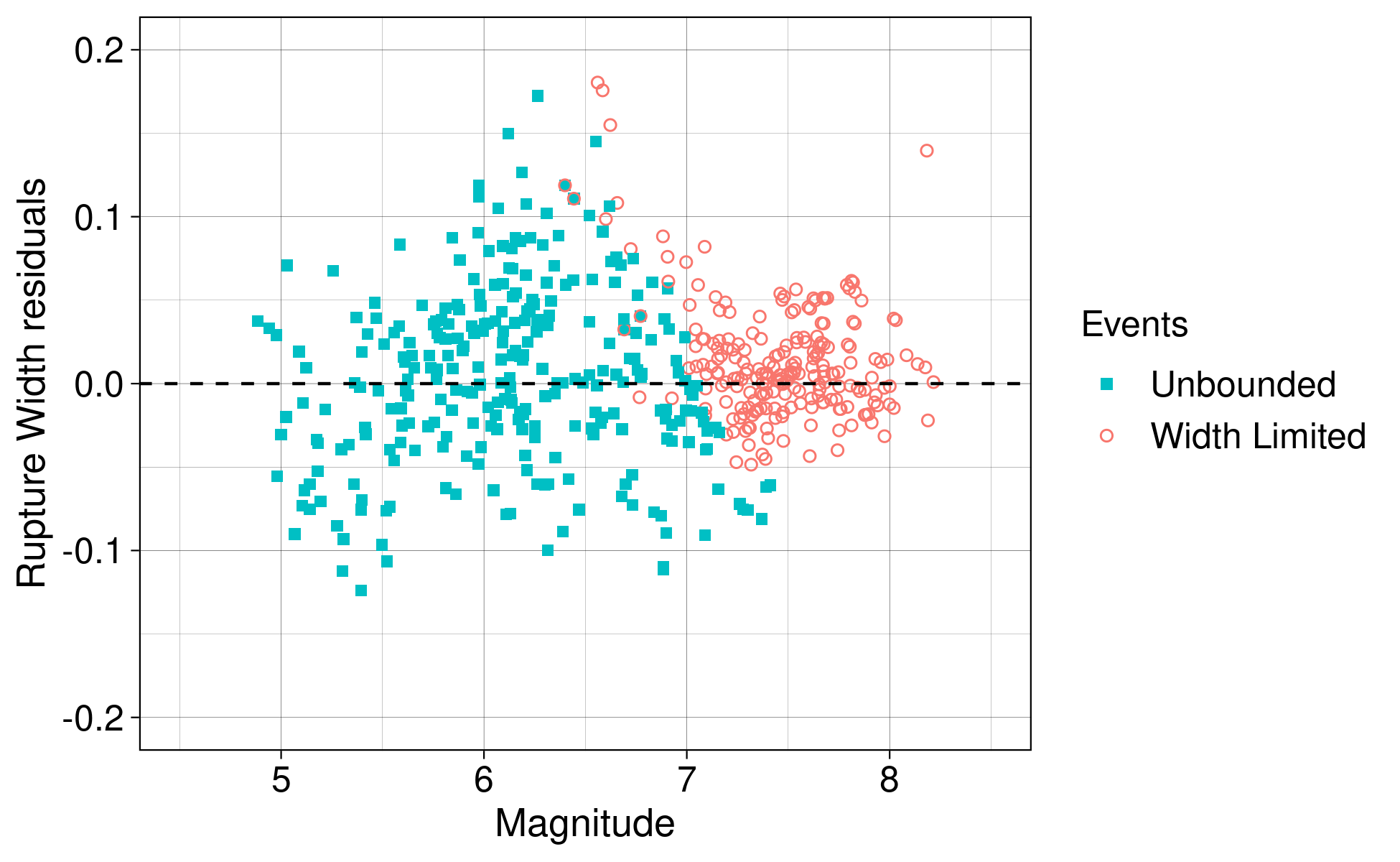}
    \caption{Residuals of Magnitude - Rupture Width model.}
    \label{fig:model_W_residuals}
\end{figure}

The effective rupture widths, estimated from the SCEC simulations, were used to determine the  width model. 
A linear functional form with a plateau for the upper limit is proposed:
\begin{equation} \label{eq:model_W_fun}
    \log_{10}(W) = \min(\beta_1 + \beta_2 \mathbf{M} + \epsilon_W, \log_{10}(W_{lim}))
\end{equation}
in which the log of the rupture width scales linearly with magnitude until it reaches the fault width, where it remains constant. 
The coefficient $\beta_1$ is the model intercept, $\beta_2$ controls the magnitude scaling, and $\epsilon_W$ is the aleatory term that is modeled with a normal distribution with a zero mean and $\sigma_W$ standard deviation ($\epsilon_W \sim \mathcal{N}(0, \sigma_W)$) 
The model coefficients were estimated with a linear regression; the best estimates and standard error values of the model coefficients are provided in Table \ref{tab:model_W_coeffs}. 
Figure \ref{fig:model_W_prediction} compares the scaling of the proposed model with the SCEC simulations, and Figure \ref{fig:model_W_residuals} presents the regression residuals, where it can be seen that the proposed model captures the scaling of the SCEC simulations and the transition between unbounded and width-limited ruptures. 
Any differences between the rupture width scaling of the empirical data and SCEC simulations will be mapped into the aleatory variability of the $SRL$ model described in the next section. 

\begin{table}[htbp!]
	\caption{Rupture width model coefficients}
	\centering
	\label{tab:model_W_coeffs}
	\begin{tabular}{| l | c | c | c |}
	    \hline 
	    Coefficient     & $\beta_1$ & $\beta_2$ & $\sigma_W$ \\
	    \hline
	    Estimate        & $-1.1602$ & $0.3395$ & $0.0544$	\\
        Standard error  & $ 0.0325$	& $0.0053$ & $-$        \\
        \hline
	 \end{tabular}
\end{table}

\subsection{Surface Rupture Length Modeling} \label{sec:model_SRL}

In developing the $SRL$ model, the candidate functional forms are derived in \hyperref[sec:model_SRL_fun_from]{Section: Functional Form Derivation}, and the 
model coefficients are estimated in \hyperref[sec:model_SRL_reg]{Section: Functional Form Derivation}  along with a discussion on the selection of the preferred model.
 
\subsubsection{Functional Form Derivation} \label{sec:model_SRL_fun_from}

Starting with the definition of the seismic moment ($M_0 = \mu~\bar{D}~L~W$), and substituting it into the definition of the moment magnitude ($\mathbf{M} = 2/3\log_{10}(M_0) - 10.7$, \cite{Kanamori1977}), a relationship between $\mathbf{M}$, average rupture slip ($\bar{D}$), subsurface rupture length ($L$), subsurface rupture width ($W$), and crust stiffness ($\mu$) is obtained:
\begin{equation}
    \mathbf{M} = \frac{2}{3}\log_{10}(\mu) + \frac{2}{3}\log_{10}(\bar{D}) + \frac{2}{3}\log_{10}(L) + \frac{2}{3}\log_{10}(W) - 10.7
\end{equation}
Treating $\mu$ as model constant and combining it with the $-10.7$ factor, the previous equation reduces to:
\begin{equation} \label{eq:model_SRL_fun_init}
    \mathbf{M} = \frac{2}{3}\log_{10}(\bar{D}) + \frac{2}{3}\log_{10}(L) + \frac{2}{3}\log_{10}(W) + c
\end{equation}
Here and in the remaining of this paper, $c$ with no subscript is used to indicate any arbitrary constant, not a specific model coefficient. 

Two end-member scaling relationships for average displacement often discussed in the literature are the L model ($\bar{D} \sim L$), in which average displacement is proportional to rupture length, and the W model ($\bar{D} \sim W$), in which average displacement is proportional to rupture width \citep{Scholz1982, Romanowicz1992, Romanowicz1993, Romanowicz1994, Scholz1994, Bodin1996, Pegler1996, Scholz1997, Scholz1998, Hanks2002, Leonard2010}. More recent studies suggest that slip is in a comprising mode between L and W models \citep{Mai2000, Bodin1996}. 
\cite{Leonard2010} proposed the $\bar{D} \sim \sqrt{A}$ model, in which the average displacement scales proportionally to the square root of the rupture area. 
Here, we investigated all three scaling relationships in building the $SRL \sim \mathbf{M}$ model, as well as two composite scaling laws that follow $\bar{D} \sim \sqrt{A}$ for the unbounded ruptures and transition to $\bar{D} \sim L$ and $\bar{D} \sim W$ for width-limited events, respectively.
In particular, the $SRL$ relationship based on $\bar{D} \sim \sqrt{A}$ scaling is denoted model 1, the relationship based on $\bar{D} \sim W$ scaling is denoted model 2, and the relationship based on $\bar{D} \sim L$ scaling is denoted model 3.
The scaling relationship which starts with  $\bar{D} \sim \sqrt{A}$ and transitions to $\bar{D} \sim W$ is denoted model 2\textprime~while the scaling relationship which transitions to $\bar{D} \sim L$ is denoted model 3\textprime.

By substituting the $\bar{D} \sim \sqrt{A}$ scaling in Equation \ref{eq:model_SRL_fun_init}, the following scaling relationship between $\mathbf{M}$, $L$, and $W$ is obtained for model 1:
\begin{equation} \label{eq:model_SRL_opt1_init}
    \mathbf{M} = \log_{10}(L) + \log_{10}(W) + c
\end{equation}
combining Equation \ref{eq:model_SRL_fun_init} with the $\bar{D} \sim {W}$ scaling, the model 2 scaling relationship becomes:
\begin{equation} \label{eq:model_SRL_opt2_init}
    \mathbf{M} = \frac{2}{3}\log_{10}(L) + \frac{4}{3}\log_{10}(W) + c
\end{equation}
while, combining Equation \ref{eq:model_SRL_fun_init} with the $\bar{D} \sim {L}$ model, model 3 scaling relationship becomes:
\begin{equation} \label{eq:model_SRL_opt3_init}
    \mathbf{M} = \frac{4}{3}\log_{10}(L) + \frac{2}{3}\log_{10}(W) + c
\end{equation}

Adopting the rupture width scaling relationship from \hyperref[sec:model_W]{Section: Rupture Width Modeling}, assuming the surface to sub-surface rupture length scaling is of the form: $log_{10}(SRL) = c_1 \log_{10}(L) + c$, and solving for $log_{10}(SRL)$, model 1 transforms to:
\begin{equation}  \label{eq:model_SRL_opt1_inter}
    \log_{10}(SRL) = c_1 \left( \mathbf{M} - \min(\beta_1 + \beta_2 \mathbf{M}, \log_{10}(W_{lim})) \right) + c 
\end{equation}
model 2 transforms to:
\begin{equation} \label{eq:model_SRL_opt2_inter}
    \log_{10}(SRL) = c_1 \left( \frac{3}{2} \mathbf{M} - 2 \min(\beta_1 + \beta_2 \mathbf{M}, \log_{10}(W_{lim})) \right) + c
\end{equation}
and model 3 transforms to:
\begin{equation} \label{eq:model_SRL_opt3_inter}
    \log_{10}(SRL) = c_1 \left( \frac{3}{4} \mathbf{M} - \frac{1}{2} \min(\beta_1 + \beta_2 \mathbf{M}, \log_{10}(W_{lim})) \right) + c
\end{equation}

Considering Equations \ref{eq:model_SRL_opt1_inter} and \ref{eq:model_SRL_opt2_inter}, the functional form the composite model 2\textprime~is:
\begin{equation} 
    \log_{10}(SRL) = 
    \begin{cases}
        c_1 \left( \mathbf{M} - \min(\beta_1 + \beta_2 \mathbf{M}, \log_{10}(W_{lim})) + \delta_1 \right) + c                  & \mathbf{M} \le \mathbf{M}_{lim} \\
        c_1 \left( \frac{3}{2} \mathbf{M} - 2 \min(\beta_1 + \beta_2 \mathbf{M}, \log_{10}(W_{lim})) \right) + c    & \mathbf{M} > \mathbf{M}_{lim} \\
    \end{cases}
\end{equation}
where $\mathbf{M}_{lim}$ is the magnitude at which the transition from unbounded to width-limited events occurs. 
The term $d_1$ is introduced to ensure a continuous transition between the two branches. 
Using Equation \ref{eq:model_W_fun}, $\mathbf{M}_{lim}$ is calculated as: $\mathbf{M}_{lim}= 1/\beta_2 ( \log_{10}(W_{lim}) - \beta_1)$.
The transition from unbounded to width-limited events occurs at the $(\mathbf{M}_{lim}, W_{lim})$ coordinate pair. Setting the two branches equal to each other at the transition point and solving for $d_1$, we obtain:
\begin{equation}
    d_1 = -\frac{1}{2} \mathbf{M}_{lim} + \log_{10}(W_{lim}) 
\end{equation}
With the previous equation, the two branches of model 2\textprime~can be combined into one equation as:
\begin{equation} 
    \log_{10}(SRL) = c_1 \left( \frac{3}{2} \mathbf{M} - \frac{1}{2} \min(\mathbf{M},\mathbf{M}_{lim}) - \min(\beta_1 + \beta_2 \mathbf{M}, \log_{10}(W_{lim})) \right) + c
\end{equation}

Similarly, cconsidering Equations \ref{eq:model_SRL_opt1_inter} and \ref{eq:model_SRL_opt3_inter}, the functional form the composite model 3\textprime~is:
\begin{equation} 
    \log_{10}(SRL) = 
    \begin{cases}
        c_1 \left( \mathbf{M} - \min(\beta_1 + \beta_2 \mathbf{M}, \log_{10}(W_{lim})) + \delta_2 \right) + c                  & \mathbf{M} \le \mathbf{M}_{lim} \\
        c_1 \left( \frac{3}{4} \mathbf{M} - \frac{1}{2} \min(\beta_1 + \beta_2 \mathbf{M}, \log_{10}(W_{lim})) \right) + c    & \mathbf{M} > \mathbf{M}_{lim} \\
    \end{cases}
\end{equation}
Following a similar derivation to  model 2\textprime, the functional form of model 3\textprime~can be expressed as:
\begin{equation} 
    \log_{10}(SRL) = c_1 \left( \frac{3}{4} \mathbf{M} + \frac{1}{4} \min(\mathbf{M},\mathbf{M}_{lim}) - \min(\beta_1 + \beta_2 \mathbf{M}, \log_{10}(W_{lim})) \right) + c
\end{equation}

The style of faulting is implicitly included in the previous scaling relationships through $W_{lim}$; Normal and Reverse faults have shallower dip angles compared to Strike-Slip faults leading to larger fault widths for the same seismological zone thickness. 
However, an additive term for dip-slip faults ($F_D$) is also added in the final functional forms to account for differences in the static stress drop between the different styles of faulting. 
In preliminary regressions, separate additive terms for Normal and Reverse faults were also evaluated but were rejected to favor model simplicity as both Reverse, and Normal faults had similar additive term values. 
The static stress drop is defined as:
\begin{equation}
    \Delta \sigma = C \mu \frac{\bar{D}}{L_C}
\end{equation}
where $L_C$ is the characteristic length, and $C$ is a constant which is a function of the rupture geometry. 
In log scale, a shift in the static stress drop will be accompanied by a shift in the average displacement ($\log(\bar{D}) = \log(\Delta \sigma) + c$), which, in turn, will shift the intercept of the $SLR \sim \mathbf{M}$ relationship.  
Thus, the final functional form for model 1 becomes:
\begin{equation} \label{eq:model_SRL_opt1_final}
\begin{aligned}
    \log_{10}(SRL) &= c_0 + c_1 \left( \mathbf{M} - \min(\beta_1 + \beta_2 \mathbf{M}, \log_{10}(W_{lim})) \right) + c_2 F_D + + \epsilon_{SRL} \\
    &= c_0 + c_1 X_1 + c_2 F_D + \epsilon_{SRL}
\end{aligned}
\end{equation}
which can be express as a linear model with $X_1 = \left( \mathbf{M} - \min(\beta_1 + \beta_2 \mathbf{M}, \log_{10}(W_{lim})) \right)$. 
The functional form for model 2 is:
\begin{equation}
\begin{aligned}
    \log_{10}(SRL) &= c_0 + c_1 \left( \frac{3}{2} \mathbf{M} - 2 \min \left( \beta_1 + \beta_2 \mathbf{M}, \log_{10}(W_{lim}) \right) \right) + c_2 F_D + \epsilon_{SRL} \\
    &= c_0 + c_1 X_2 + c_2 F_D + c_3 + \epsilon_{SRL}
\end{aligned}
\end{equation}
with $X_2 = \left( 3/2 \mathbf{M} - 2 \min \left( \beta_1 + \beta_2 \mathbf{M}, \log_{10}(W_{lim}) \right) \right) $ to be expressed as a linear model.
The functional form for model 3 is:
\begin{equation}
\begin{aligned}
    \log_{10}(SRL) &= c_0 + c_1 \left( \frac{3}{4} \mathbf{M} - \frac{1}{2} \min \left( \beta_1 + \beta_2 \mathbf{M}, \log_{10}(W_{lim}) \right) \right) + c_2 F_D + \epsilon_{SRL} \\
    &= c_0 + c_1 X_3 + c_2 F_D + c_3 + \epsilon_{SRL}
\end{aligned}
\end{equation}
with $X_3 = \left( 3/4 \mathbf{M} - 1/2 \min \left( \beta_1 + \beta_2 \mathbf{M}, \log_{10}(W_{lim}) \right) \right) $.
Model 2\textprime~final functional form is:
\begin{equation}
\begin{aligned}
    \log_{10}(SRL) =& c_0 + c_1 \left( \frac{3}{2} \mathbf{M} - \frac{1}{2} \min(\mathbf{M},\mathbf{M}_{lim}) - \min \left( \beta_1 + \beta_2 \mathbf{M}, \log_{10}(W_{lim}) \right) \right) \\
     &+ c_2 F_D + \epsilon_{SRL} \\
    =& c_0 + c_1 X'_2 + c_2 F_D + c_3 + \epsilon_{SRL}
\end{aligned}
\end{equation}
with $X'_2 = \left( \frac{3}{2} \mathbf{M} - \frac{1}{2} \min(\mathbf{M},\mathbf{M}_{lim}) - \min \left( \beta_1 + \beta_2 \mathbf{M}, \log_{10}(W_{lim}) \right) \right)$, and model 3\textprime~final functional form is:
\begin{equation}
\begin{aligned}
    \log_{10}(SRL) =& c_0 + c_1 \left( \frac{3}{4} \mathbf{M} + \frac{1}{4} \min(\mathbf{M},\mathbf{M}_{lim}) - \min \left( \beta_1 + \beta_2 \mathbf{M}, \log_{10}(W_{lim}) \right) \right) \\
     &+ c_2 F_D + \epsilon_{SRL} \\
    =& c_0 + c_1 X'_3 + c_2 F_D + c_3 + \epsilon_{SRL}
\end{aligned}
\end{equation}
with $X'_3 = \left( \frac{3}{4} \mathbf{M} + \frac{1}{4} \min(\mathbf{M},\mathbf{M}_{lim}) - \min \left( \beta_1 + \beta_2 \mathbf{M}, \log_{10}(W_{lim}) \right) \right)$.

Lastly, a Wells and Coppersmith type model was also evaluated, hereafter referred to as model 0, to investigate the impact of width-limited ruptures on the current state of practice models. 
The functional form for model 0 is:
\begin{equation}
    \log_{10}(SRL) = c_0 + c_1 \mathbf{M} + c_2 F_D + \epsilon_{SRL}
\end{equation}
In all previous equations $c_0$ is the model intercept, $c_1$ controls the magnitude scaling, $c_2$ captures the median shift between strike-slip and dip-slip events ($F_D$ is zero for strike-slip and one for reverse and normal faults), and $\epsilon_{SRL}$ is the aleatory term. 
More information on the modeling of the aleatory variability is provided at the end of \hyperref[sec:model_SRL_reg]{Subsection: Model Regression}.

\subsubsection{Model Regression} \label{sec:model_SRL_reg}

All candidate models were estimated using a maximum likelihood linear regression and empirical dataset.
Table \ref{tab:model_L_logL_AIC_candiate} provides the log-likelihood ($\mathcal{L}$) and Akaike Information Criterion (AIC) for the different models; a higher $\mathcal{L}$ and lower AIC indicate a better model fit to the data. 
Table \ref{tab:model_L_coeffs_candiate} summarizes the best estimates and standard errors of the model coefficients.

From a statistical perspective, candidate models 1 to 3 have similar good performance, model 2\textprime~and 3\textprime~have poorer performance, and model 0 has the worst performance showcasing the limitations of a purely empirical model. Models 1 to 3\textprime~have a break in the magnitude scaling when the width of the rupture reaches the width of the seismogenic zone, allowing them to fit the data better. 

\begin{table}[htbp!]
	\caption{Log-likelihood ($\mathcal{L}$) and Akaike Information Criterion (AIC) of the candidate rupture length models}
	\centering
	\label{tab:model_L_logL_AIC_candiate}
	\begin{tabular}{|l|c|c|}
	    \hline
        Model               & $\mathcal{L}$            & AIC                       \\ \hline
	    Model 0             & $3.24$                   & $1.52$                    \\ \hline
        Model 1             & $13.16$                  & $-18.33$                  \\ \hline
        Model 2             & $12.95$                  & $-17.91$                  \\ \hline
        Model 3             & $11.40$                  & $-14.80$                  \\ \hline
        Model 2\textprime   & $9.06$                   & $-10.11$                  \\ \hline
        Model 3\textprime   & $9.01$                   & $-10.30$                  \\ \hline
    \end{tabular}
\end{table}

\begin{table}[htbp!]
	\caption{Rupture length candidate model coefficients}
	\centering
	\label{tab:model_L_coeffs_candiate}
	\begin{tabular}{|l|l|c|c|c|c|}
	    \hline
	    Model                               &  Coefficient      & $c_0$         & $c_1$     & $c_2$     & $\sigma_{SRL-unb}$  \\ 
	    \hline
	    \multirow{2}{*}{Model 0}            & Estimate          & $-2.8684$     & $0.6481$  & $-0.1708$ & $0.2386$  \\
	                                        & Standard error    & $0.2192$      & $0.0315$  & $0.0496$  & $-$       \\ \hline
	    \multirow{2}{*}{Model 1}            & Estimate          & $-3.8512$     & $0.9470$  & $-0.1439$ & $0.2527$  \\
	                                        & Standard error    & $0.2267$      & $0.0377$  & $0.0463$  & $-$       \\ \hline
	    \multirow{2}{*}{Model 2}            & Estimate          & $-4.1834$     & $0.7163$  & $-0.1313$ & $0.2532$  \\
	                                        & Standard error    & $0.2403$      & $0.0286$  & $0.0466$  & $-$       \\ \hline
        \multirow{2}{*}{Model 3}            & Estimate          & $-3.5568$     & $1.1220$  & $-0.1564$ & $0.2564$  \\
	                                        & Standard error    & $0.2188$      & $0.0454$  & $0.0468$  & $-$       \\ \hline
	    \multirow{2}{*}{Model 2\textprime}  & Estimate          & $-3.1307$     & $0.8111$  & $-0.1209$ & $0.615$   \\
	                                        & Standard error    & $0.2064$      & $0.0336$  & $0.0484$  & $-$       \\ \hline
        \multirow{2}{*}{Model 3\textprime}  & Estimate          & $-4.1787$     & $1.0114$  & $-0.1672$ & $0.2615$  \\
	                                        & Standard error    & $0.2497$      & $0.0419$  & $0.0419$  & $-$       \\
        \hline
	 \end{tabular}
\end{table}

Based on seismological theory, model 1 to 3\textprime~should have unit slopes with respect to the magnitude scaling terms (e.g. $X_1$, $X_2$, $X_3$), if the average displacement in the empirical data follows the assumed scaling law and the subsurface rupture length ($L$) was the response variable. 
With $SRL$ as the response variable, a magnitude scaling term greater than one implies that the surface to subsurface rupture length ratio increases with magnitude, while a magnitude scaling term less than one implies the opposite. 
Intuitively, a larger than one magnitude scaling term is expected, as for large events, a bigger part of the rupture is expected to reach the surface. 
Considering that model 1 exhibited the best predictive performance, but the magnitude scaling term was slightly less than one, the final preferred model uses model 1 functional form (Equation \ref{eq:model_SRL_opt1_final}) with a fixed unit slope. 
The statistical measures for the goodness of fit and coefficients of the preferred model are summarized in Table \ref{tab:model_L_logL_AIC_preferred} and \ref{tab:model_L_coeffs_preferred}, respectively.
The proposed model shows good predictive performance (third largest log-likelihood and smallest AIC out of all the considered models) as well as a reasonable extrapolation behavior due to the imposed constraints.

\begin{table}[htbp!]
	\caption{Log-likelihood ($\mathcal{L}$) and Akaike Information Criterion (AIC) of the preferred model}
	\centering
	\label{tab:model_L_logL_AIC_preferred}
	\begin{tabular}{|l|c|c|}
	    \hline
        Model                & $\mathcal{L}$             & AIC                       \\ \hline
        Model 1, fixed slope & \multirow{2}{*}{$12.16$}  & \multirow{2}{*}{$-18.33$} \\ 
        (preferred model)    &                           &                           \\ \hline
    \end{tabular}
\end{table}

\begin{table}[htbp!]
	\caption{Rupture length preferred model coefficients}
	\centering
	\label{tab:model_L_coeffs_preferred}
	\begin{tabular}{|l|l|c|c|c|c|}
	    \hline
	    Model                  &  Coefficient      & $c_0$         & $c_1$     & $c_2$     & $\sigma_{SRL-unb}$  \\ 
	    \hline
	    Model 1, fixed slope   & Estimate          & $-4.1673$     & $1$       & $-0.1207$ & $0.2537$  \\
	    (preferred model)      & Standard error    & $0.0251$      & $-$       & $0.0434$  & $-$       \\ \hline
	 \end{tabular}
\end{table}


The findings of the regression analyses favor a $\sim \sqrt{A}$ scaling for the average displacement compared to $\sim W$ and $\sim L$ scaling relationships consistent with \cite{Hanks2002}. 
The SCEC simulations show some extra support of $\sim \sqrt{A}$ scaling. While a limited width ($15 km$ and $19 km$) is set for all simulations, the smooth transition near the limited width and deeper scattering width for larger earthquakes indicate that a fixed limited width is not suitable for the general scaling. As shown in simulations of \cite{Wang2022}, the deep penetration of large earthquakes keeps increasing average displacements with the rupture length, that deviates from a purely W-based scaling and approaches the $\sqrt{A}$ scaling. In addition, \cite{Thingbaijam2017} observed $D \sim A^{0.429 \pm 0.134}$ for reverse events, $D \sim A^{0.858 \pm 0.214}$ for normal events and $D \sim A^{0.597 \pm 0.112}$ for strike-slip events, which supports the preference of the $\sqrt{A}$ over W scaling.

\begin{figure}
    \centering
    \includegraphics[height=0.5\textwidth]{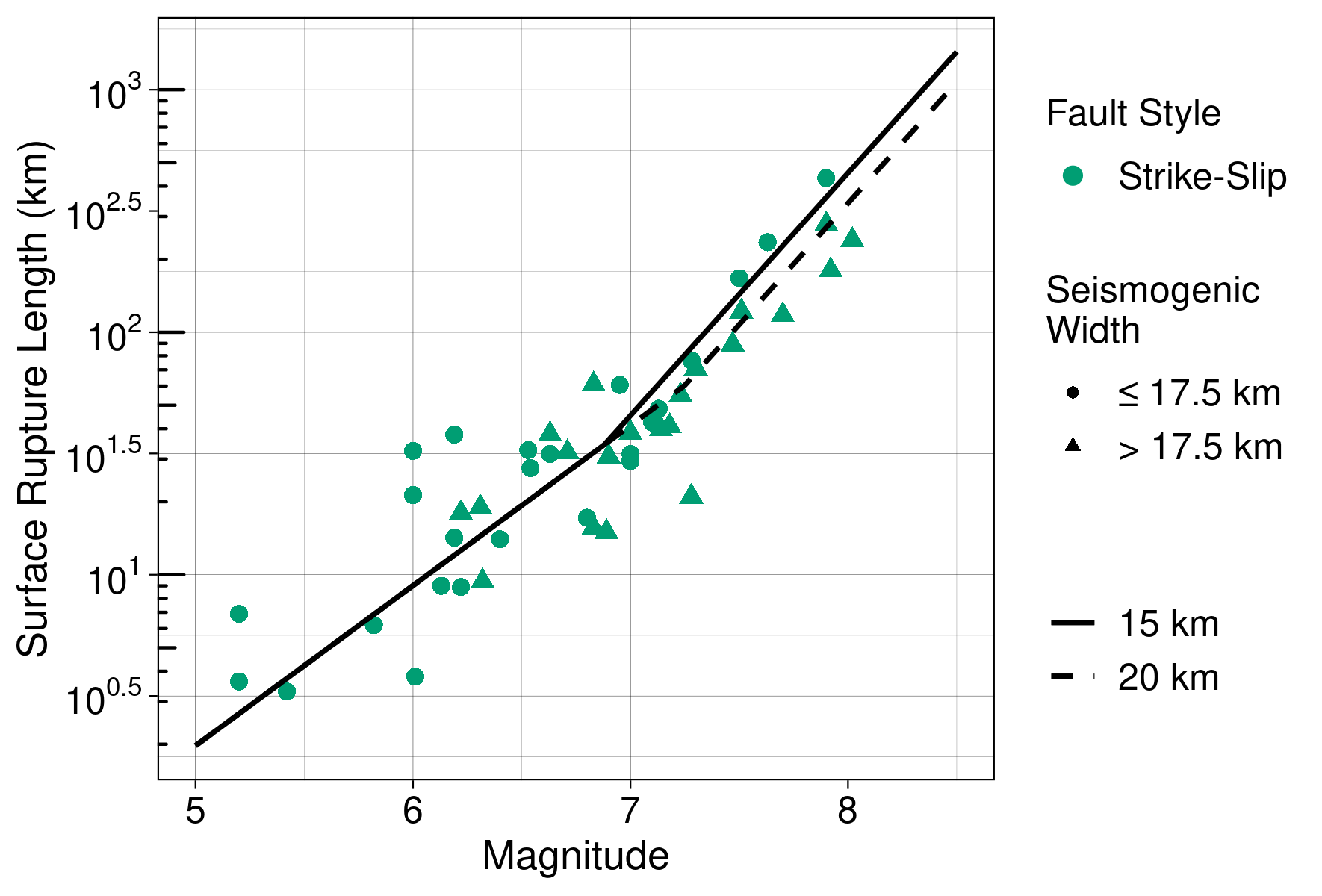}
    \caption{Magnitude scaling of model 1 with fixed slope (preferred model) for Strike-slip events with a $15 km$ thickness of seismogenic zone, shown with the solid line, and a $20 km$ thickness of seismogenic zone, shown with the dashed line.
    Circular markers correspond to Strike-slip events on seismological zones less than $17.5~km$ thick, and triangular markers correspond to Strike-slip events on seismological zones between $17.5$ and $25~km$ thick.}
    \label{fig:model_pref_SLR_mag_slr}
\end{figure}

Figure \ref{fig:model_pref_SLR_mag_slr} presents the median magnitude scaling for the preferred model for Strike-slip events for a seismological zone thickness equal to $15$ and $20~km$ against the empirical data for the same slip type and seismological zone thickness up to $25~km$.
For the $15~km$ thick zone, the magnitude break occurs at $\mathbf{M}~6.9$.
For the $20~km$ crust, a larger rupture area is needed for the event to become width limited; thus, the magnitude break occurs closer to $\mathbf{M}~7.3$. 
The empirical data show consistent trends; at small magnitudes, there is no systematic difference between events in thin and thick parts of the seismogenic zones, while at larger magnitudes, there is a positive shift for events in thin parts of the seismogenic zone compared to events on thicker seismogenic zones. 

\begin{figure}
    \centering
    \includegraphics[height=0.5\textwidth]{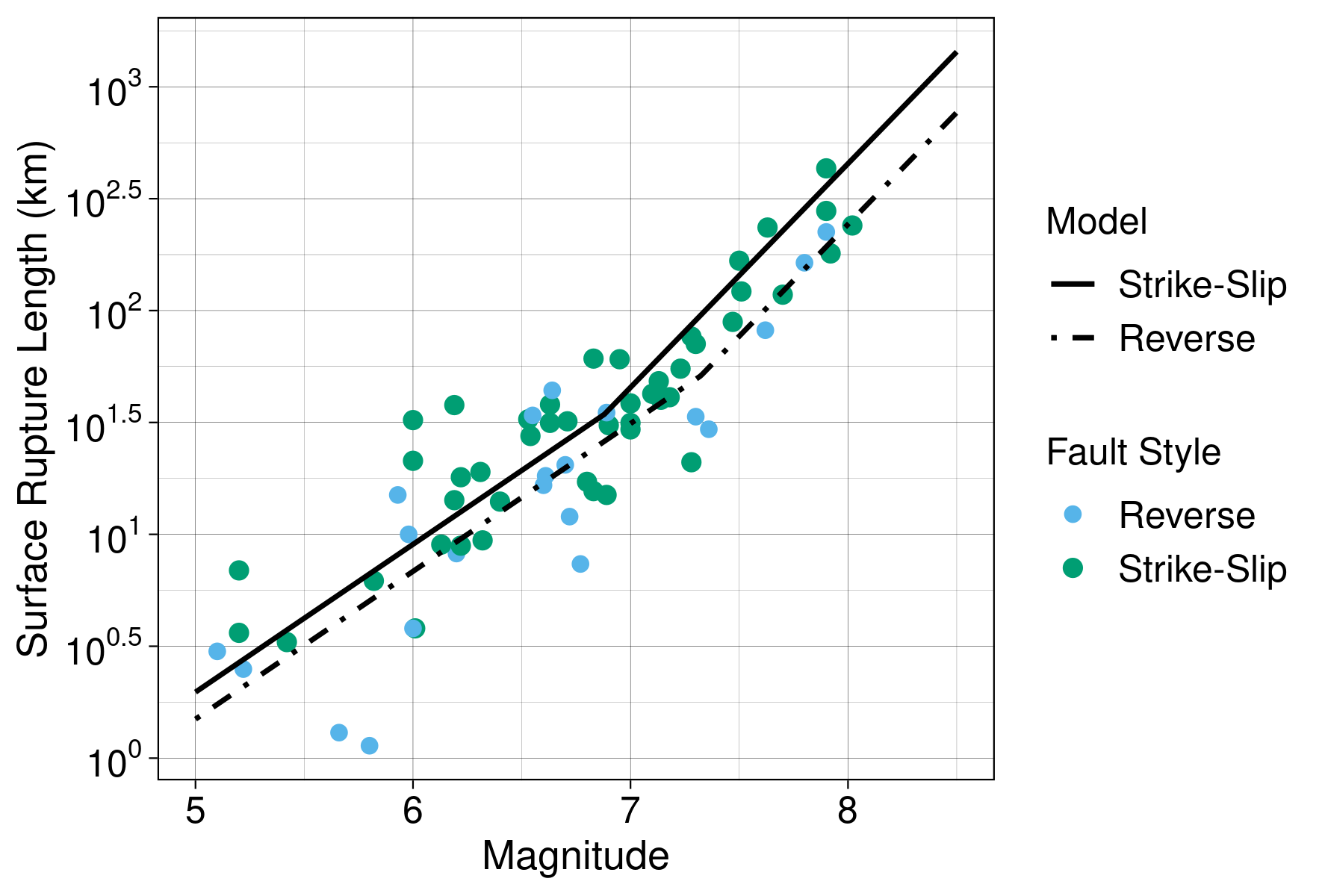}
    \caption{Magnitude scaling of model 1 with fixed slope (preferred model) for strike-slip events, shown with the solid line, and reverse-slip events, shown with the dotted-dashed line. Both cases were evaluated for a seismological zone thickness of $15 km$.
    The empirical data for Strike-slip events on seismogenic zones up to $25~km$ are also depicted; green markers correspond to Strike-slip events and blue markers correspond to Reverse events.}
    \label{fig:model_pref_SLR_mag_slr_FD}
\end{figure}

Figure \ref{fig:model_pref_SLR_mag_slr_FD} shows the preferred model scaling for Strike-slip and Reverse-slip events for a seismological zone thickness equal to $15 km$. 
The vertical offset between the two models is introduced by the $F_D$ term, and the difference in the magnitude break is caused by the different dip angles between Strike-slip and Reverse faults. 
Although the same seismological thickness was assumed for both scaling relationships, the shallower dip angle of Reverse faults allows for a wider fault width which can accommodate larger events before they become width limited.  
The magnitude break for the Strike-slip events occurs at $\mathbf{M}~6.9$ and for reverse events at $\mathbf{M}~7.3$.
Consistent trends are also observed in the empirical data.

\begin{figure}[htbp!]
    \centering
    \begin{subfigure}[t]{0.48\textwidth}
        \caption{}
        \includegraphics[height = 0.78\textwidth]{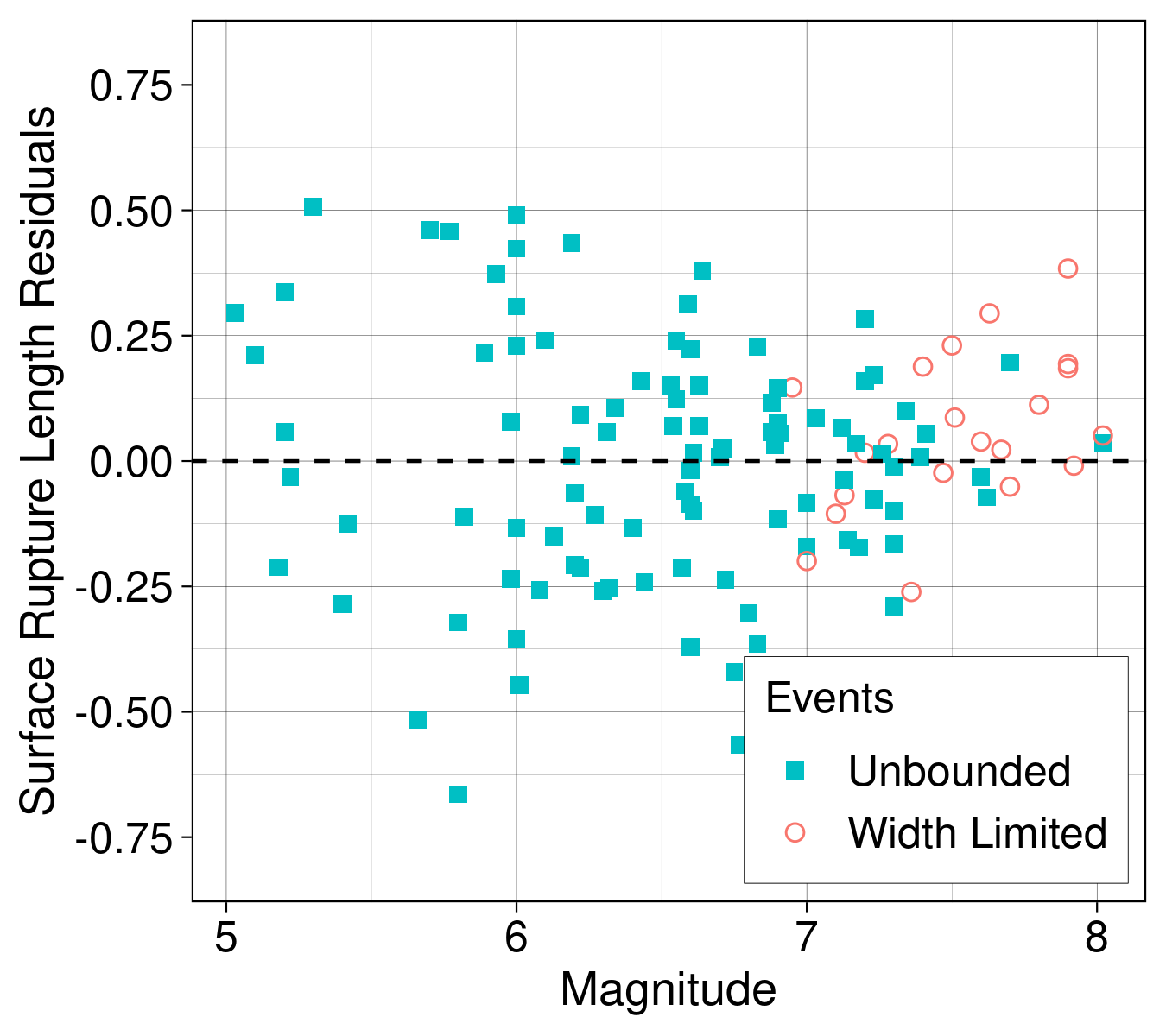}
        \label{fig:model0_SRL_residuals_mag}
    \end{subfigure} 
    \begin{subfigure}[t]{0.48\textwidth}
        \caption{}
        \includegraphics[height = 0.78\textwidth]{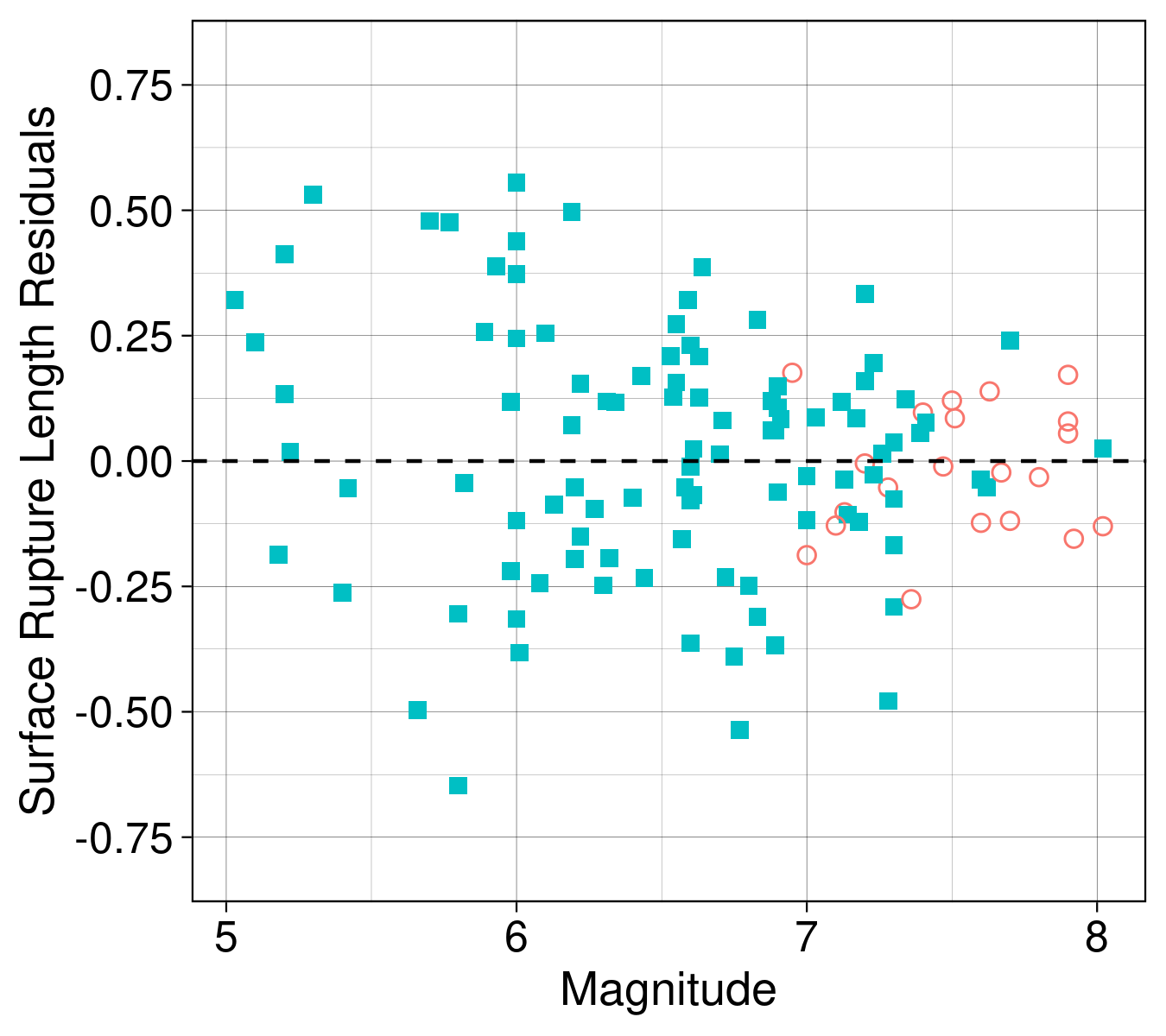}
        \label{fig:model1_fxd_SRL_residuals_mag}
    \end{subfigure}  
    \caption{Comparison of regression residuals versus magnitude for (a) model 0 and (b) model 1 with fixed slope. }
    \label{fig:model_SRL_residuals_mag}
\end{figure}

Figure \ref{fig:model_SRL_residuals_mag} compares the regression residuals from model 0 and model 1 with a fixed slope.
Both models have similar zero-mean-centered residuals for unbounded ruptures.
However, subfigure \ref{fig:model0_SRL_residuals_mag} shows a positive bias in the model 0 residuals for width-limited ruptures, whereas in subfigure \ref{fig:model1_fxd_SRL_residuals_mag} the model 1 residuals for width-limited ruptures are centered closer to zero. 
This comparison illustrates the advantage of a seismological-theory-based model; using a quadratic functional form would address the positive bias for the residuals but provides no basis for its existence.
Relating the change in magnitude scaling to the finite thickness of the seismogenic zone increases the confidence in an extrapolation behavior that is scientifically defensible.


\begin{figure}
    \centering
    \includegraphics[height=0.5\textwidth]{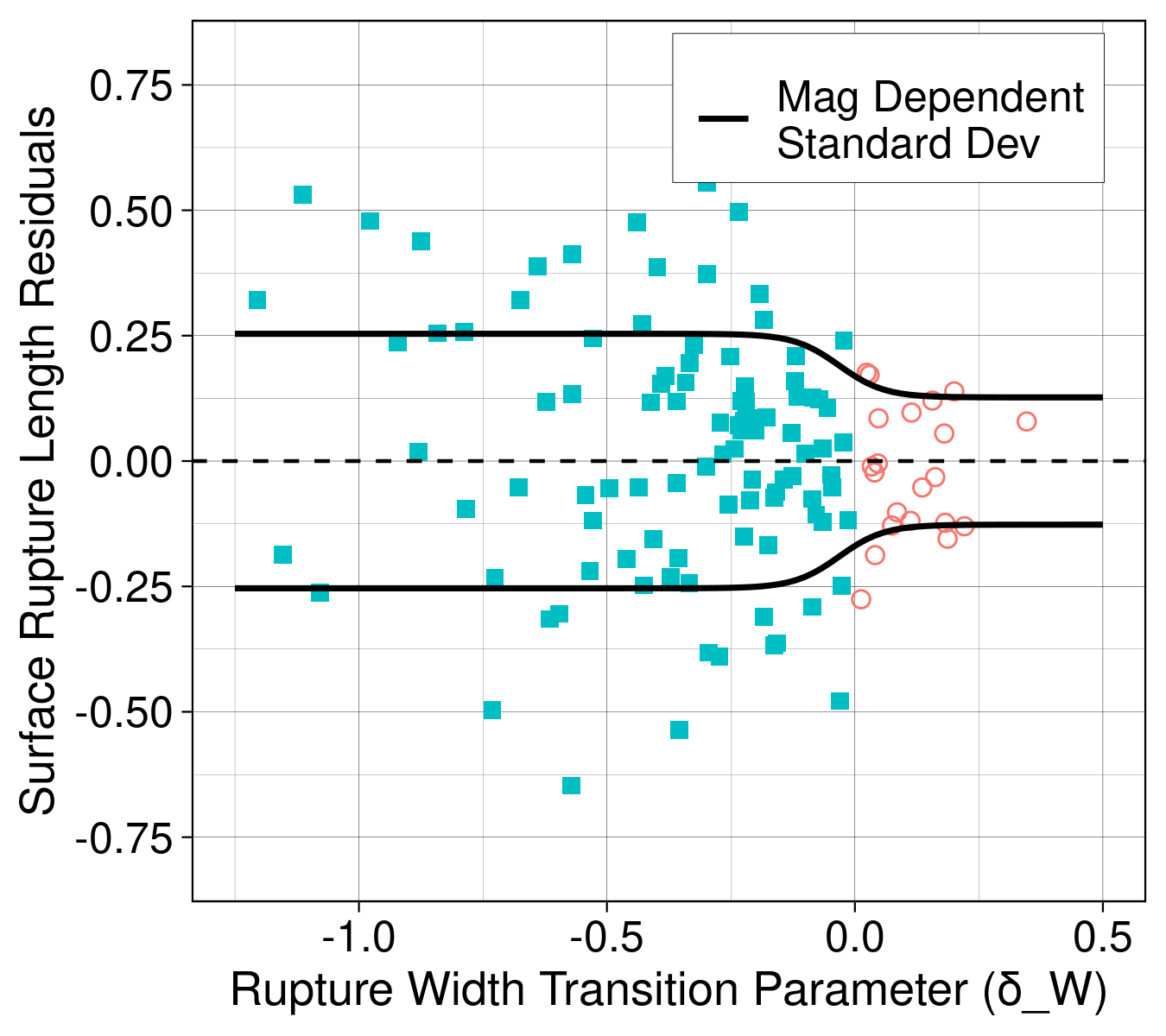}
    \caption{The proposed aleatory standard deviation for model 1, shown with the solid lines, and regression residuals versus the rupture width transition parameter ($\delta_W$).}
    \label{fig:model1_fxd_SRL_residuals_W}
\end{figure}

Preliminary regressions showed a magnitude-dependent aleatory variability that is by a factor of two smaller for width-limited ruptures as compared to unbounded ruptures.
A rupture width transition parameter ($\delta_W$) is introduced, defined as:
\begin{equation}
    \delta_W = \log_{10}(W_{unb}) - \log_{10}(W_{lim}) =  \beta_1 \mathbf{M} + \beta_2 - \log_{10}(W_{lim})
\end{equation}
where $W_{unb}$ is the theoretical rupture width for an infinite thickness seismological zone estimated from the first leg of rupture width model ($\log_{10}(W_{unb}) = \beta_1 \mathbf{M} + \beta_2$); a negative $\delta_W$ corresponds to an unbounded rupture, a positive $\delta_W$ corresponds to a width-limited rupture, and the transition phase occurs around zero. 
Using the parameter $\delta_W$, a heteroscedastic standard deviation model for the aleatory variability is proposed ($\epsilon_{SLR} \sim \mathcal{N}(0, \sigma_{SLR}(\delta_W))$) that gradually shifts from unbounded to width-limited ruptures based on a sigmoid function:
\begin{equation}
    \sigma_{SLR}(\delta_W) = \frac{\sigma_{SRL-unb}} {1 + S( 10\ln(9) ~ \delta_W)}
\end{equation}
where $\sigma_{SLR}$ is the magnitude-dependent standard deviation, and $\sigma_{SRL-unb}$ is the standard deviation for unbounded ruptures reported in Table \ref{tab:model_L_coeffs_preferred}. 
The sigmoid function is defined as ($S(x) = 1/(1+\exp(-x))$). 
Figure \ref{fig:model1_fxd_SRL_residuals_W} presents the regression residuals and proposed standard deviation versus the parameter $\delta_W$ for the preferred model.
The factor $10 \times \ln(9)$ is used so that $80\%$ of the standard-deviation change occurs between $\delta_W = -0.1$ and $0.1$, which is consistent with the empirical observations. 

\section{Comparisons}

A series of comparisons with available models was performed to ensure the reasonableness of the proposed relationship. 

\begin{figure}
    \centering
    \includegraphics[height=0.5\textwidth]{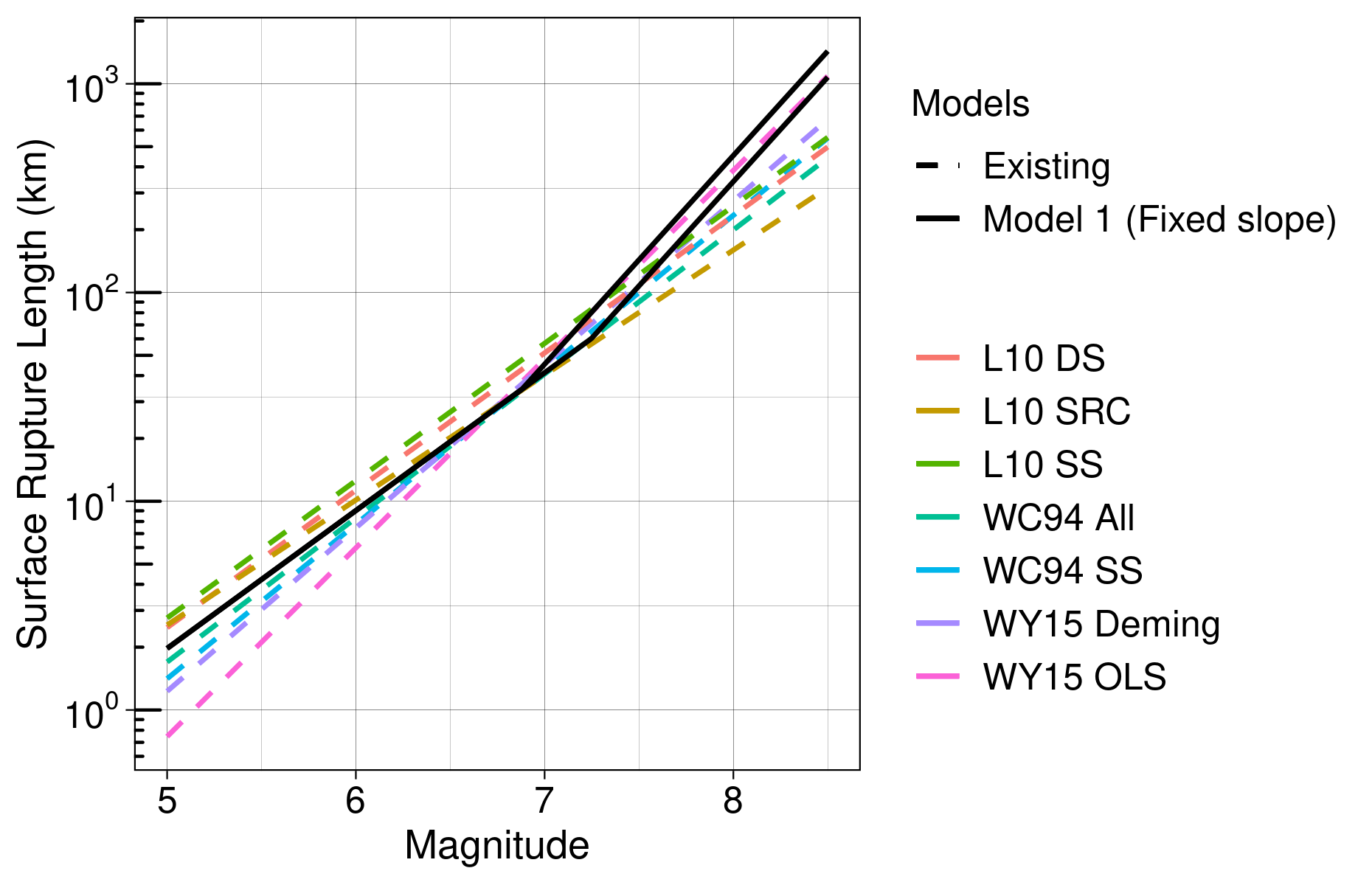}
    \caption{Comparison of the preferred model (model 1 with fixed slope) for strike-slip events on a $15$ and $20km$ thick crust with existing $SLR$ relationships.}
    \label{fig:model1_fxd_SRL_comparison}
\end{figure}

Figure \ref{fig:model1_fxd_SRL_comparison} compares the preferred model with existing $SLR$ relationships;
``WC94 All'' and ``WC94 SS'' correspond to the \cite{Wells1994} $SRL \sim \mathbf{M}$ scaling relationships for all and Strike-slip events, ``WY15 OLS'' and ``WY15 Deming'' correspond to the \cite{Wells2015} $SRL \sim \mathbf{M}$ scaling relationships using ordinary-least-squares and errors-in-variables regression models. 
``L10 DS'', ``L10 SS'', and ``L10 SRC'' correspond to the \cite{Leonard2010} $SRL \sim \mathbf{M}$ scaling relationships for Dip-slip and Strike-slip in active crustal regions, and stable-continental events.

The proposed model is in overall agreement with the existing relationships over different magnitude ranges. 
It is in good agreement with the \cite{Leonard2010} scaling relationship for small to moderate-size events and in good agreement with the \cite{Wells1994} and \cite{Wells2015} scaling relationships at large-size events. 
A potential reason for the agreement at different magnitude ranges is that the dataset used by \cite{Leonard2010} was primarily composed of events ranging from $\mathbf{M}~5$ to $7.2$, while datasets used by \cite{Wells1994} and \cite{Wells2015} was primarily composed of events ranging from $\mathbf{M}~6$ to $7.8$.

\section{Range of Applicability}

A $\mathbf{M}~5$ lower limit is recommended as it is the smallest magnitude in the regression dataset. 
For the upper limit, the largest event in the dataset was $\mathbf{M}~8.1$.
However, the authors believe that seismological constraints will help the extrapolation up to $\mathbf{M}~8.5$, albeit with increased epistemic uncertainty.

\section{Discussion and Conclusions}

The proposed $SRL$ model captures the change in magnitude scaling between unbounded and width-limited ruptures through the use of seismological constraints and dynamic fault rupture simulations. 
Seismological theory was used to derive candidate scaling relationships between the moment magnitude ($M$), subsurface length ($L$), and rupture width ($W$). 
The dynamic rupture simulations constrained the $W \sim \mathbf{M}$ scaling.
The empirical data were used to decide between the alternative scaling relationships for average displacement, to model the difference between $L$ and $SRL$ scaling, and to capture the aleatory variability in $SRL$ scaling.
The empirical data supports a $\sqrt{A}$-type scaling for average displacement compared to $W$-type and $L$-type scaling.

Compared to a simple linear regression between $SLR$ and $\mathbf{M}$, the proposed functional form provides a better fit to empirical datasets. 
Both the linear and the proposed functional forms provide similar fits for unbounded ruptures, but the proposed relationship provides a better fit for width-limited events. 
A quadratic functional form could have been used to resolve the positive bias at large magnitudes, but it would lack a seismological basis limiting the confidence for its extrapolation. 
A width transition parameter, $\delta_W$, is proposed to capture the difference in aleatory variability between unbounded and width-limited ruptures. 
The parameter $\delta_W$ corresponds to the log of the ratio of the theoretical rupture width for an infinitely thick seismogenic and the actual fault width.  
A comparison with other existing models was performed.
The proposed model is in good agreement with \cite{Leonard2010} for small to moderate events and in good agreement with \cite{Wells1994} and \cite{Wells2015} for moderate to large events. 
It is believed this difference is caused by the different magnitude ranges in the datasets of the existing models.

Future studies should evaluate the effect of the thickness of the seimogenic zone using fault-specific information. 
Additionally, a fully-consistent scaling model for magnitude, rupture area, subsurface rupture length and width, and surface rupture length for unbounded and width-limited ruptures using a more comprehensive dataset following the presented framework is encouraged.

\section*{Acknowledgments}

This work was supported by the California Energy Commission, California Department of Transportation, Pacific Gas and Electric Company, and Southern California Earthquake Center (SCEC), SCEC Contribution \#XXXX. SCEC is funded by the National Science Foundation (NSF) and U.S. Geological Survey (USGS) through cooperative agreements with the University of Southern California (USC). Any opinions, findings, and conclusions or recommendations expressed in this material are those of the authors and do not necessarily reflect those of the sponsoring agencies. An award of computer time was provided by the INCITE program and this research used resources of the Argonne Leadership Computing Facility, which is a DOE Office of Science User Facility supported under Contract DE-AC02-06CH11357. This research was also partially computed on Frontera computing project at the Texas Advanced Computing Center through an allocation made possible by National Science Foundation award OAC-1818253. 

The authors would also like to thank Stephane Baize for providing the \cite{Baize2020} dataset and Alexandra Sarmiento for the insightful discussions during the development of this model.

\section*{Data and Resources}

The regression code and regression datasets are provided in: \url{https://github.com/NHR3-UCLA/LWABC22_SRL_model}
The statistical regressions were performed using the computer software R and stats package \citep{R2022}.
The open-source software package Support Operator Rupture Dynamics (SORD) can be downloaded from \url{https://github.com/wangyf/sordw3}.

\bibliographystyle{abbrvnat}
\bibliography{references_mendely_GL, references_other}

\end{document}